\documentclass[iop]{emulateapj}

\def\gtorder{\mathrel{\raise.3ex\hbox{$>$}\mkern-14mu
             \lower0.6ex\hbox{$\sim$}}} 
\def\ltorder{\mathrel{\raise.3ex\hbox{$<$}\mkern-14mu
             \lower0.6ex\hbox{$\sim$}}}
\def\ltsima{$\; \buildrel < \over \sim \;$}
\def\simlt{\lower.5ex\hbox{\ltsima}}
\def\gtsima{$\; \buildrel > \over \sim \;$}
\def\simgt{\lower.5ex\hbox{\gtsima}} 

\shorttitle{CCCP SN II Light Curves}
\shortauthors{Arcavi et al.}

\begin{document} 


\title{Caltech Core-Collapse Project (CCCP) Observations of Type II Supernovae: Evidence for Three Distinct Photometric Subtypes}


\author{Iair~Arcavi\altaffilmark{1}$^{,}$\altaffilmark{2},
Avishay~Gal-Yam\altaffilmark{1},
S. Bradley Cenko\altaffilmark{3},
Derek B. Fox\altaffilmark{4},
Douglas C. Leonard\altaffilmark{5},
Dae-Sik Moon\altaffilmark{6},
David J. Sand\altaffilmark{7}$^{,}$\altaffilmark{8},
Alicia M. Soderberg\altaffilmark{9},
Michael Kiewe\altaffilmark{10},
Ofer Yaron\altaffilmark{1},
Adam B. Becker\altaffilmark{3},
Raphael Scheps\altaffilmark{11},
Gali Birenbaum\altaffilmark{12},
Daniel Chamudot\altaffilmark{13}
and Jonathan Zhou\altaffilmark{14}
}

\altaffiltext{1}{Department of Particle Physics and Astrophysics, The Weizmann Institute of Science, Rehovot 76100, Israel}
\altaffiltext{2}{iair.arcavi@weizmann.ac.il}
\altaffiltext{3}{Department of Astronomy, University of California, Berkeley, CA 94720-3411, USA}
\altaffiltext{4}{Department of Astronomy and Astrophysics, Pennsylvania State University, University Park, PA 16802, USA}
\altaffiltext{5}{Department of Astronomy, San Diego State University, San Diego, CA 92182, USA}
\altaffiltext{6}{Department of Astronomy and Astrophysics, University of Toronto, Toronto, ON M5S 3H4, Canada}
\altaffiltext{7}{Las Cumbres Observatory Global Telescope Network, Santa Barbara, CA 93117, USA}
\altaffiltext{8}{Department of Physics, University of California, Santa Barbara, CA 93106, USA}
\altaffiltext{9}{Harvard-Smithsonian Center for Astrophysics, Cambridge, MA 02138, USA}
\altaffiltext{10}{Department of Physics, University of Wisconsin, Madison, WI 53706, USA}
\altaffiltext{11}{King's College, University of Cambridge, Cambridge, CB2 1ST, UK}
\altaffiltext{12}{12 Amos St, Ramat Chen, Ramat Gan, 52233, Israel}
\altaffiltext{13}{20 Chen St, Petach Tikvah, 49520, Israel}
\altaffiltext{14}{101 Dunster St., Box 398, Cambridge, MA 02138, USA}



\newpage

\begin{abstract} 

We present $R$-Band light curves of Type~II supernovae (SNe) from the Caltech Core Collapse Project (CCCP). With the exception of interacting (Type~IIn) SNe and rare events with long rise times, we find that most light curve shapes belong to one of three distinct classes: plateau, slowly declining and rapidly declining events. The last class is composed solely of Type~IIb SNe which present similar light curve shapes to those of SNe Ib, suggesting, perhaps, similar progenitor channels. We do not find any intermediate light curves, implying that these subclasses are unlikely to reflect variance of continuous parameters, but rather might result from physically distinct progenitor systems, strengthening the suggestion of a binary origin for at least some stripped SNe. We find a large plateau luminosity range for SNe~IIP, while the plateau lengths seem rather uniform at approximately $100$ days. As analysis of additional CCCP data goes on and larger samples are collected, demographic studies of core collapse SNe will likely continue to provide new constraints on progenitor scenarios.

\end{abstract} 




\section{Introduction} 

Type~II supernovae (SNe) are widely recognized as the end stages of massive H-rich stars and represent the bulk of observed core collapse SNe (see Filippenko 1997 for a review of SN classifications). Several sub-types of Type~II SNe have been observed. Those showing a plateau in their light curve are known as Type~IIP events, while those showing a linear decline from peak magnitude are classified as IIL. A third class of events, Type~IIb, characterized by its spectral rather than its photometric properties, develops prominent He features at late times. Finally, Type~IIn SNe display narrow lines in their spectra, indicative of interaction between the SN ejecta and a dense circum stellar medium. 

Red supergiants (RSGs) have been directly identified as the progenitors of Type~IIP SNe for SN2003gd (Van Dyk et al. 2003; Smartt et al. 2004), SN2004A (Hendry et al. 2006), SN2005cs (Maund et al. 2005; Li et al. 2006), SN2008bk (Mattila et al. 2008; Van Dyk et al. 2012) and SN2009md (Fraser et al. 2011); see Smartt 2009 for a review. Such stars have thick hydrogen envelopes that are ionized by the explosion shock wave. As the shocked envelope expands and cools, it recombines, releasing radiation at a roughly constant rate, thus producing a plateau in the light curve (e.g. Popov 1993; Kasen \& Woosley 2009). It follows that SNe IIL might be the explosions of stars with less massive H envelopes that can not support a plateau in their light curve. SN IIb progenitors, then, would contain an even smaller H mass.

However, if SNe IIP-IIL-IIb progenitors represent merely a sequence of decreasing H-envelope mass, one would expect the properties of these SNe to behave as a continuum. Specifically, a gradual transition in light curve shape should be observed when examining a homogeneous sample of events. 

The Caltech Core Collapse Project (CCCP; Gal-yam et al. 2007) is a large observational survey which made use of the robotic 60-inch (P60; Cenko et al. 2006) and Hale 200-inch telescopes at Palomar Observatory to obtain optical $BVRI$ photometry and spectroscopy of 48 nearby core collapse SNe. By providing a fair sample of core collapse events with well defined selection criteria and uniform, high quality optical observations, CCCP allows to study core collapse SNe as a population rather than as individual events. 

Light curves of Type~Ib/c SNe from CCCP have been presented and analyzed by Drout et al. (2011). Type~IIn CCCP events are treated by Kiewe et al. (2012). Here we present photometry of 21 non-interacting Type~II SNe with well observed light curves collected through CCCP. We present $R$-Band data for most of the events to simplify the comparison of their light curve shapes. A more detailed multi-color analysis will be presented in a forthcoming paper.

\section{Photometry}

Our light curves are produced using image subtraction with respect to P60 reference imaging obtained approximately one year or later after explosion. We employ the CPM method (Gal-Yam et al. 2008a) for PSF matching using the mkdifflc routine (Gal-Yam et al. 2004) implemented in IRAF\footnote{IRAF (Image Reduction and Analysis Facility) is distributed by the National Optical Astronomy Observatories, which are operated by AURA, Inc., under cooperative agreement with the National Science Foundation.}. Our photometry is calibrated to Sloan Digital Sky Survey (SDSS) stars near the SN - their magnitudes converted to the Johnson-Cousins systems using the equations of Jordi et al. (2005). For objects outside the SDSS footprint we used Landolt standards observed at the same night as the SN field for calibration. Our measurements are presented in the natural system (see Kiewe et al. 2012 for more details) with typical photometric errors of $\sim0.1$ magnitude. The photometry is corrected for Galactic extinction using the Schlegel et al. (1998) maps retrieved via the NASA/IPAC Extragalactic Database (NED). The distance moduli to the SN host galaxies are taken from NED, if available, or calculated from spectroscopic redshifts assuming a cosmological model with $H_0 = 70\,\textrm{km\,s}^{-1}\,\textrm{Mpc}^{-1}$, $\Omega_{m}=0.3$ and $\Omega_{\Lambda}=0.7$ otherwise. We adopt a distance modulus of 29.62 for SN 2005cs, based on the distance estimate of Vink{\'o} et al. (2012; hereafter V12).

Due to incomplete data for three of the events, we use photometry published in the literature for them. The light curve of SN2004fx is taken from Hamuy et al. (2006), that of SN2005ay from Gal-Yam et al. (2008b), and that of SN2005cs from Pastorello et al. (2009).

For each SN, we constrain the explosion date to a window between the last non-detection and first detection. For a few events, this window is wider than $14$ days, but the supernova was detected before peak brightness or the first spectrum taken displayed a blue continuum, known to be indicative of a young event (e.g., Gal-Yam et al. 2011). We treat these cases as if a non-detection existed $14$ days prior to the first detection. The explosion date window widths are noted in Table \ref{subtypes}. Two events with poorly constrained explosion dates did not display blue featureless spectra upon first detection. These are SN2005au and SN2005bw (both SNe IIP) and we include them only to study absolute plateau luminosities. 

\section{Results and Discussion}

\begin{figure*}
\includegraphics[width=18cm]{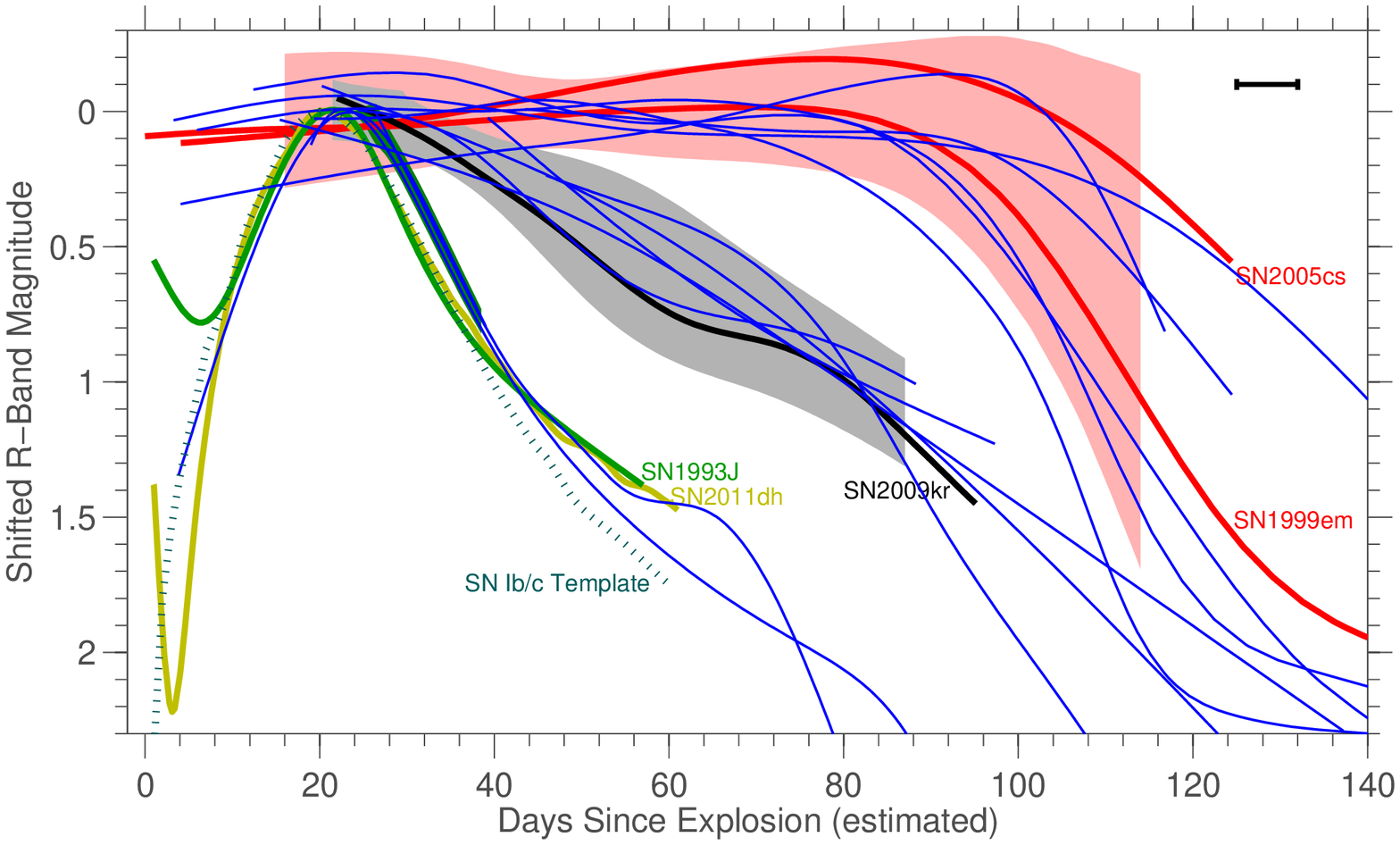}
\includegraphics[width=18cm]{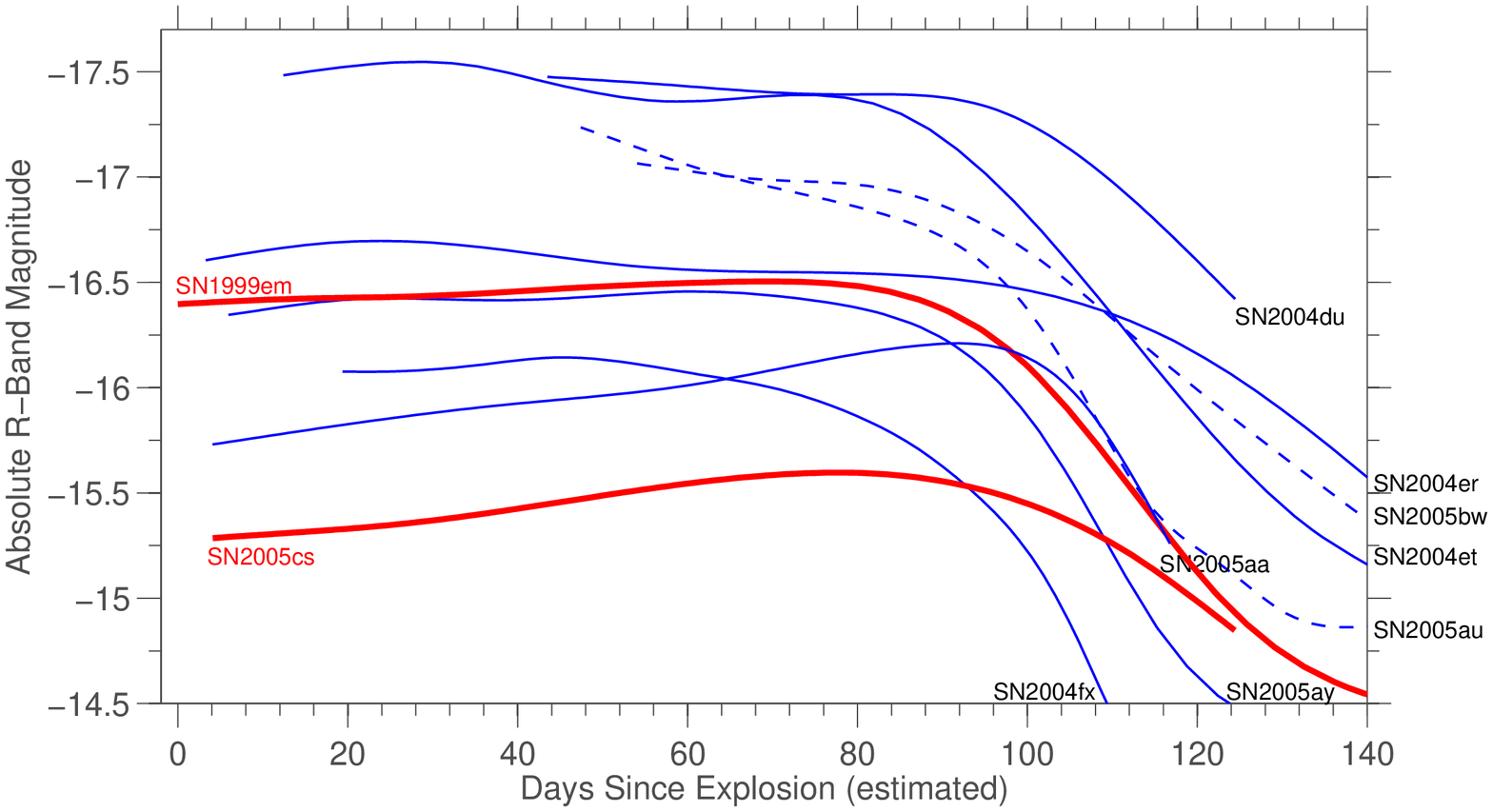}
\caption{Top Panel: $R$-band light curves of 15 Type~II SNe from CCCP (excluding the four events presented in Figure \ref{pecs}), normalized in peak magnitude (SN2004fx data taken from Hamuy et al. 2006; SN2005ay data taken from Gal-Yam et al. 2008b; SN2005cs data taken from Pastorello et al. 2009). A clear subdivision into three distinct subtypes is apparent: plateau, slowly declining and rapidly declining SNe (the latter consisting only of SNe IIb). Reference SNe are shown for comparison (SN1999em from Leonard et al. 2002; SN2009kr from Fraser et al. 2010, found to be a member of the IIL subclass as claimed by Elias-Rosa et al. 2010; SN1993J from Richmond et al. 1994; SN2011dh from Arcavi et al. 2011). We also overplot the SN Ib/c template derived by Drout et al. (2011). The data have been interpolated with spline fits (except for SN2005by, where a polynomial fit provided a better trace to the data). The shaded regions denote the average light curve $\pm2\sigma$ of each subclass. The maximal 7-day uncertainty in determining the explosion times is illustrated by the interval in the top right corner.
Bottom Panel: $R$-band light curves of 9 Type~IIP SNe from CCCP with respect to their estimated explosion time (except for SN2005au and SN2005bw, marked by dashed lines, for which the explosion date is not known to good accuracy). The light curve of SN2004fx is taken from Hamuy et al. (2006), that of SN2005ay from Gal-Yam et al. (2008b) and that of SN2005cs from Pastorello et al. (2009). SN1999em (Leonard et al. 2002) is shown for comparison. A spread in plateau luminosities is apparent while plateau lengths seem to converge around 100 days. Spline fits were applied to the data.}
\label{alllcs}
\end{figure*}

We plot the $R$-Band light curves of $15$ Type~II events normalized to peak magnitude in Figure \ref{alllcs} (top panel). Rather than forming a continuum, we find that the light curves group into three distinct sub-classes: plateau, slowly declining (1-2 Mag/100 days) and initially rapidly declining (5-6 Mag/100 days) events (see also Table \ref{subtypes}). We note that the three rapidly declining events are all Type~IIb and that they display similar light curve shapes to those of Type~Ib/c SNe (Drout et al. 2011). We perform a Kolmogorov-Smirnov test and find that the probability that the measured $\Delta\textrm{M}15_{R}$ values for the rapid and slow decline groups are drawn from a single underlying distribution is $2\%$.

\begin{figure*}
\includegraphics[width=18cm]{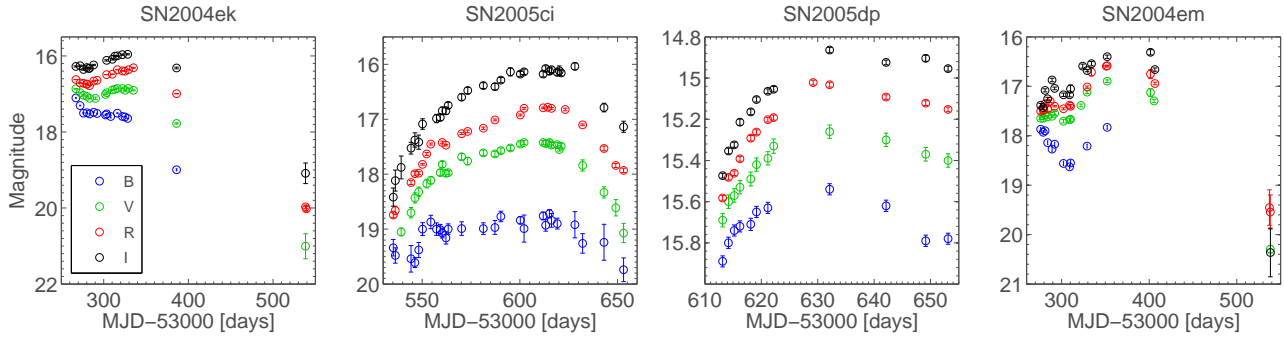}
\caption{$BVRI$ light curves of the CCCP events not included in Figure \ref{alllcs}. Three events (SN2004ek, SN2005ci, SN2005dp) show long rise times (possibly associated with blue supergiant explosions; Kleiser et al. 2011; Pastorello et al. 2012), while one peculiar event (SN2004em) changes behavior from flat to rising around three weeks after explosion.}
\label{pecs}
\end{figure*}

Three events (SN2004ek, SN2005ci and SN2005dp; Figure \ref{pecs}) display prolonged rising periods in their light curves. They do not show signs of interaction in their spectra and may be explosions of compact blue supergiant progenitors (Kleiser et al. 2011; Pastorello et al. 2012), as demonstrated directly in the case of SN 1987A (see Arnett et al. 1989 for a review). 

Finally, one event (SN2004em; Figure \ref{pecs}) displays a very peculiar photometric behavior. For the first few weeks it is similar to a Type~IIP SN, while around day $25$ it suddenly changes behavior to resemble a SN 1987A-like event.

The full photometric dataset is available online through WISeREP\footnote{http://www.weizmann.ac.il/astrophysics/wiserep} (Yaron \& Gal-Yam 2012).

\begin{figure*}
\includegraphics[width=18cm]{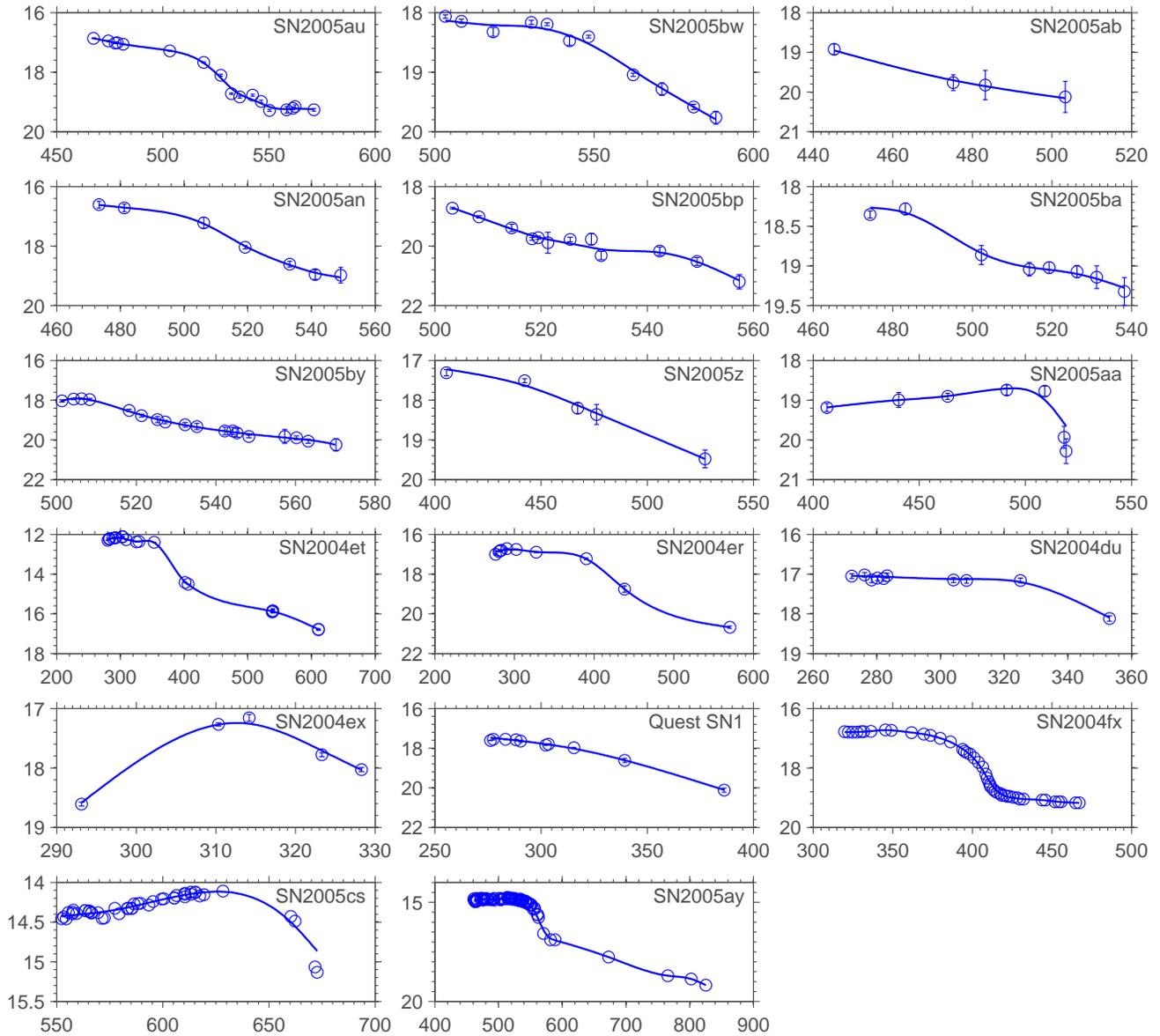}
\caption{$R$-Band photometric data of the CCCP events included in Figure \ref{alllcs} together with the spline fits shown in that figure. The x-axis is MJD-53000 in days, and the y-axis is apparent magnitude. The light curve of SN2004fx is taken from Hamuy et al. (2006), that of SN2005ay from Gal-Yam et al. (2008b) and that of SN2005cs from Pastorello et al. (2009).}
\label{spline}
\end{figure*}

\begin{deluxetable*}{llllllll}
\tablecolumns{8}
\tablewidth{0pt}
\tablecaption{Subdivision of the CCCP Type~II SN light curves analyzed. For the declining SNe, the peak magnitudes and the $\Delta\textrm{M}15_{R}$ parameter (denoting the magnitude drop at 15 days after peak in the $R$-band) are derived from the smooth fits to the light curves (limits are stated when the rise to peak is not detected in the data). For the plateau events, the average luminosity during the first 50 days is taken as the plateau magnitude (one $\sigma$ shown in parentheses). The explosion dates assumed in Fig. \ref{alllcs} are noted (explosion date window widths shown in parentheses).}
\tablehead{
\colhead{SN} &
\colhead{Type} &
\colhead{Peak/Plateau Mag} & 
\colhead{$\Delta\textrm{M}15_{R}$} &
\colhead{DM} &
\colhead{DM Source} &
\colhead{Explosion MJD} &
\colhead{Reference}    
\\
}
\startdata

SN2004du & Plateau & $-17.464$ ($0.055$) & & $34.26$ & NED & $53229$ ($2$) & IAUC 8387 \\
SN2004er & Plateau & $-16.669$ ($0.077$) & & $33.39$ & NED & $53273$ ($2$) & IAUC 8412 \\
SN2004et & Plateau & $-17.482$ ($0.088$) & & $28.80$ & NED & $53268$ ($4$) & IAUC 8413 \\
SN2004fx\footnote{photometry from Hamuy et al. 2006} & Plateau & $-16.087$ ($0.038$) & & $32.60$ & NED & $53300$ ($3$) & IAUC 8431 \\
SN2005aa & Plateau & $-15.824$ ($0.134$) & & $34.83$ & Redshift & $53402$ ($5$) & IAUC 8476 \\
SN2005au & Plateau & $-17.069$ ($0.141$) & & $34.07$ & NED & & IAUC 8496 \\
SN2005ay\footnote{photometry from Gal-Yam et al. 2008b} & Plateau & $-16.447$ ($0.119$) & & $31.21$ & NED & $53452$ ($4$) & IAUC 8500/2 \\
SN2005bw & Plateau & $-16.945$ ($0.150$) & & $35.15$ & Redshift & & CBET 147 \\
SN2005cs\footnote{photometry from Pastorello et al. 2009} & Plateau & $-15.040$ ($0.079$) & & $29.62$ & V12 & $53548$ ($2$) & IAUC 8553\\
\hline
SN2005Z & Slow Decline & $<-17.5$ & $0.135$ & $34.61$ & Redshift & $53391$ ($5$) & IAUC 8476 \\
SN2005ab & Slow Decline & $<-15.2$ & $0.405$ & $34.14$ & Redshift & $53406$ ($7$) & IAUC 8478 \\
SN2005an & Slow Decline & $<-17.0$ & $0.170$ & $33.39$ & Redshift & $53430$ ($7$) & CBET 113 \\
SN2005ba & Slow Decline & $-17.4$ & $0.202$ & $35.60$ & Redshift & $53450$ ($7$) & IAUC 8503 \\
Quest SN1 & Slow Decline & $-16.5$ & $0.181$ & $33.63$ & Redshift & $53256$ ($7$) & \\
\hline
SN2004ex & Rapid Decline & $-17.2$ & $0.751$ & $34.41$ & Redshift & $53289$ ($1$) & IAUC 8418 \\
SN2005bp & Rapid Decline & $<-16.8$ & $0.926$ & $35.41$ & Redshift & $53477$ ($7$) & IAUC 8515 \\
SN2005by & Rapid Decline & $-17.5$ & $0.750$ & $35.39$ & Redshift & $53482$ ($6$) & IAUC 8523 \\
\hline
SN2004ek & Prolonged Rise & $-18.2$ & & $34.38$ & NED & & IAUC 8405 \\
SN2005ci & Prolonged Rise & $-16.1$ & & $32.80$ & NED & & IAUC 8541 \\
SN2005dp & Prolonged Rise & $-17.6$ & & $32.57$ & NED & & IAUC 8591\\
SN2004em & Peculiar & $-17.9$ & & $34.05$ & Redshift & & IAUC 8406 

\enddata
\label{subtypes}
\end{deluxetable*}

\subsection{Declining SNe}

Aside from establishing a different rate of decline for SNe IIb compared to SNe IIL, Figure \ref{alllcs} (top panel) suggests that the IIP, IIL and IIb subtypes do not span a continuum of physical parameters, such as H envelope mass. Rather, additional factors should be considered. Specifically, Type~IIb events might arise from binary systems (as suggested also by recent progenitor studies for SN1993J, Maund et al. 2004; SN2008ax, Crockett et al. 2008; SN2011dh, Arcavi et al. 2011, Van-Dyk et al. 2011). The similarity of the Type~IIb light curves to those of Type~Ib events (also seen in the Drout et al. 2011 data), in addition to the known spectral similarities at late times and the similar peak radio luminosities (Chevalier \& Soderberg 2010), suggests that these two types of events might come from similar progenitor systems.

\subsection{Plateau SNe}

The $R$-Band light curves of the Type~IIP SNe, on an absolute magnitude scale, can be seen in Figure \ref{alllcs} (bottom panel). We find a wide range of plateau luminosities, but do not have enough statistics to test whether they form a continuous distribution or if there are two distinct underlying types (bright and faint), as previously suggested (Pastorello et al. 2004). The plateau lengths, however, seem rather uniform at $\sim100$ days (with the sole exception of SN2004fx, displaying a shorter plateau)\footnote{Note that SN2005au and SN2005bw are plotted only to show their plateau luminosity, their plateau lengths are unknown due the lack of sufficient constraints on their explosion time}. This is consistent with the plateau length scale given by Popov (1993), which assumes a constant opacity:
$$t_{p}=99\frac{\kappa_{0.34}^{1/6}M_{10}^{1/2}R_{0,500}^{1/6}}{E_{51}^{1/6}T_{\textrm{ion},5054}^{2/3}}\textrm{days}$$
(where the radius and mass are expressed in solar units). However, according to this scaling, one would expect to see also longer plateaus (up to $\sim130$ days for $17M_\odot$ progenitors). Kasen \& Woosley (2009) show that heating from radioactive decay of $^{56}\textrm{Ni}$ should further extend the duration of the plateau ($0.1M_\odot$ of $^{56}\textrm{Ni}$ extends the plateau by $\sim24\%$ in their models, without greatly affecting the plateau luminosity). We do not find any of these long duration plateaus in our sample, though a few such events have been observed (e.g. Hamuy 2002).

The scarcity of observed short plateaus and the possibly related sharp distinction between IIP and IIL light curves is evident. Such a pronounced absence of intermediate events might suggest that Type~IIL SNe are powered by a different mechanism than that associated with SNe~IIP (e.g., magnetars; Kasen \& Bildsten 2010).

\section{Summary}

We identify a subdivision of Type~II SNe into three main photometric sub-classes as well as several peculiar events among the CCCP sample. The distinct sub-class division suggests that type IIb, IIL and IIP SNe are not members of one continuous class and may result from different physical progenitor systems: Type~IIP from single RSGs, Type~IIL possibly related to magnetars and Type~IIb (and by association perhaps also Type~Ib) from interacting binaries. We do not find any Type IIP events with plateaus longer than $\sim100$ days, in contrast to theoretical expectations (but see also Hamuy 2002).

Our dataset can be used to put SNe into context via their light curves. SN2009kr, for example, clearly belongs to the IIL subclass, as claimed by Elias-Rosa et al. (2010) and does not behave photometrically like a common SN~IIP. SN2003ie, on the other hand, is not consistent with being a SN IIL, but rather fits the IIP subclass (Arcavi et al., in prep.).

We present multi-color light curves of three additional long-rising events, possibly related to the explosions of blue supergiants (Kleiser et al. 2011; Pastorello et al. 2012) and one peculiar SN which displays an abrupt change in its photometric behavior.

We plan to complete the release of CCCP data in a forthcoming paper. Incorporating multi-color light curves and spectroscopic information into the current analysis promises to shed more light on this intriguing subtype division, and consequently on the possible progenitor scenarios leading to the different SN types. 

It is clear that statistical analyses of core collapse SNe as a population are powerful tools to study the gap in the massive star - SN mapping. With several large scale transient surveys underway, additional results are expected in the near future. One such survey, the Palomar Transient Factory (PTF; Rau et al. 2009, Law et al. 2009), has discovered, classified and followed over 400 core collapse SNe to date, more than half of which are non-interacting Type~II events. These larger statistics will help test and better quantify the results presented here.

We thank Mansi M. Kasliwal and Richard S. Walters for scheduling P60 reference imaging, and Yossi Shvartzvald for obtaining the required calibration data for SN2005an. A.G. and I.A. acknowledge support by the Israeli and German-Israeli Science Foundations, and the Lord Sieff of Brimpton fund. S.B.C.~acknowledges generous financial assistance from Gary \& Cynthia Bengier, the Richard \& Rhoda Goldman Fund, NASA/{\it Swift} grants NNX10AI21G and NNX12AD73G, the TABASGO Foundation, and US National Science Foundation (NSF) grant AST-0908886.

\end{document}